# Giant Lateral Optical Forces on Rayleigh Particles near Hyperbolic and Extremely Anisotropic Metasurfaces

N. K. Paul, D. Correas-Serrano, J. S. Gomez-Diaz*

Department of Electrical and Computer Engineering, University of California Davis, One Shields Avenue, Kemper Hall 2039, Davis, CA 95616, USA

*jsgomez@ucdavis.edu

*We report a dramatic enhancement of the lateral optical forces induced on electrically polarizable Rayleigh particles near hyperbolic and extremely anisotropic metasurfaces under simple plane wave illumination. Such enhancement is enabled by the interplay between the increased density of states provided by these structures and the out-of-plane polarization spin acquired by the particle. The resulting giant lateral forces appear over a broad frequency range and may open unprecedented venues for routing, trapping, and assembling nanoparticles.*

PACS: 32.10.Dk, 42.25.Fx, 73.20.Mf, 78.67.Wj

Light-induced forces have led to many exciting applications in nanotechnology and bioengineering by trapping, pushing, pulling, binding and manipulating nanoparticles and biological samples [1-8]. Recently, the emergence of nano-optical plasmonic configurations has been exploited to boost the strength of optical forces at the nano-scale by exciting surface plasmon polaritons (SPPs) [9-12], which are confined electromagnetic waves traveling along dielectric-metal interfaces [13]. In fact, illuminating with light an electric, non-magnetic, Rayleigh particle (with radius $r < \lambda_0/20$, being $\lambda_0$ the wavelength) located above a metallic surface like gold or silver induces optical forces on the particle that compensate the momentum of the directional SPPs excited in the scattering process, leading to a strong dependence between the induced forces and the density of states provided by the surface [14]. Such response has been further enhanced taking advantage of the quantum spin-Hall effect of light [15-17], exploiting the spin of circularly polarized incoming waves to control the excitation of directional plasmons and, in turn, the direction of the induced forces [18,19]. Over the years, a wide variety of devices have been put forward to shape these forces and merge them with other techniques such as surface-enhanced Raman spectroscopy (SERS) [20-24] or photoluminescence [25,26], aiming to select and handle nanoparticles after their individual characterization. Despite recent advances in this field [14,18,27,28], the exponential growth of nanophotonics and bioengineering is continuously imposing challenging demands to enhance the strength and control the direction of the forces induced on nanoparticles using low-intensity laser beams.

In a related context, hyperbolic metasurfaces (HMTSs) [29-38] –which are ultrathin surfaces that exhibit metallic or dielectric responses as a function of the electric field polarization– have recently gathered considerable attention from the scientific community [32-34]. These structures significantly enhance the available local density of states, support the propagation of confined surface plasmons, and have empowered an ample set of near-field functionalities such as boosting the spontaneous emission rate of emitters located nearby [32], an unusual spin-control of light [35,36], or a wavelength-dependent routing of SPPs. HMTSs can easily be realized using ultrathin nanostructured composites made of hexagonal boron nitride [31], silver [29], gold [37], graphene [35] and other 2D materials [38]. Even though the bulk version of HMTSs, hyperbolic metamaterials [39-41], have already been suggested to provide new knobs to augment optical forces [42-44], their relatively weak near-field interactions with external dipoles [45] and challenging fabrication process have hindered the use of hyperbolic structures in practical applications.

In this letter, we show that replacing standard plasmonic surfaces with hyperbolic or extremely anisotropic metasurfaces enhances the strength of the lateral optical forces induced on Rayleigh particles located nearby by several orders of magnitude. The proposed concept is illustrated in Fig. 1. Upon adequate plane wave illumination, the particle behaves as an out-of-plane circularly polarized electric dipole that excites directional SPPs on the surface thanks to the photonic spin-Hall effect [46,47] (see Fig. 1c). Due to momentum conservation, a lateral recoil force that strongly depends on the wavevector of the excited modes [14,18] is induced on the particle towards the direction opposite to the excited plasmons. Since the SPPs supported by HMTSs possess very high wavenumbers compared to isotropic surfaces, the induced recoil forces over these structures are dramatically enhanced. Fig. 1b shows that such enhancement is maximum in the very near field of the metasurface and progressively lessens as the nanoparticle moves away from it. In the following, we analytically derive the Lorentz forces acting on polarizable Rayleigh particles over anisotropic metasurfaces and describe the physical mechanisms that enable them. Then, we numerically demonstrate the presence of giant lateral optical forces over realistic HMTSs made of nanostructured silver, enabling a unique platform to route and trap nanoparticles with low-intensity laser beams.

Let us consider a non-magnetic, electrically polarizable dielectric Rayleigh particle located at a distance $z_0$ over an anisotropic metasurface defined by the diagonal conductivity tensor $\bar{\sigma} = \sigma_{xx}\hat{x}\hat{x} + \sigma_{yy}\hat{y}\hat{y}$ and illuminated by a plane wave, as shown in Fig. 1a. The optical force **F** induced on the particle can be expressed as [13,14]

$$\mathbf{F} = \frac{1}{2}\text{Re}\{\mathbf{p}^* \cdot (\nabla \otimes \mathbf{E}^{\text{loc}})\}, \tag{1}$$

where '·' and '⊗' are the dot and dyadic vector products, respectively, and '*' denotes complex conjugate. In addition, $\mathbf{E}^{\text{loc}} = \mathbf{E}^0 + \mathbf{E}^s$ is the total local electric field above the metasurface, with incident ($\mathbf{E}^i$) and reflected ($\mathbf{E}^r$) fields gathered together in the term $\mathbf{E}^0$; $\mathbf{E}^s$ is the electric field scattered by the particle, and $\mathbf{p} = \bar{\boldsymbol{\alpha}} \cdot \mathbf{E}^0$ is the electric dipole moment induced on a particle with an electric effective polarizability tensor $\bar{\boldsymbol{\alpha}}$ [48]. The optical force $\mathbf{F} = \mathbf{F}^0 + \mathbf{F}^S$ can be expressed as the superposition of the radiation pressure ($\mathbf{F}^0$) generated by the incident wave and its reflection on the surface plus the force resulting from the interaction ($\mathbf{F}^S$) of the dipolar particle with its own retarded field. Assuming an $e^{-i\omega t}$ time dependence, the lateral components of such forces can be obtained as [48]

$$F_t^0 = \frac{1}{2}k_t\left[\text{Im}\{\alpha_{xx}\}|E_x^0|^2 + \text{Im}\{\alpha_{yy}\}|E_y^0|^2 + \text{Im}\{\alpha_{zz}\}|E_z^0|^2\right], \tag{2}$$

$$F_t^S = \frac{6\pi}{c_0 k_0^2}P_{\text{rad}}^{tz}\eta_m \text{Im}\left[\frac{d}{dt}G_{tz}^S(\mathbf{r}_0, \mathbf{r}_0)\right], \tag{3}$$

where $c_0$ and $k_0$ are the speed of light in vacuum and the free-space wavenumber, respectively; $\mathbf{r}_0$ is the particle position, $t = \{x, y\}$ and $m = \{y, x\}$ (i.e., t and m axes are orthogonal), $k_t$ refers to the t component of the incident plane wave wavevector, and $G_{tz}^S(\mathbf{r}_0, \mathbf{r}_0)$ is the tz component of the scattered Green's function tensor of the system evaluated at the source position [48]. Besides, $P_{\text{rad}}^{tz} = \omega^4(|p_t|^2 + |p_z|^2)/12\pi\epsilon_0 c_0^3$ is the power radiated by the tz-component of the dipole in the absence of a metasurface, related to the strength of the induced dipole excitation [18]. Finally, $\eta_t = \left(|p_{\sigma^+}^t|^2 - |p_{\sigma^-}^t|^2\right)/\left(|p_{\sigma^+}^t|^2 + |p_{\sigma^-}^t|^2\right)$ denotes the helicity (also known as chirality factor) of the oscillating dipole as a measure of its polarization spin around the t axis [18,48]. In the expression, the subscript $\sigma^+$ ($\sigma^-$) refers to the right handed (left handed) components of the particle polarization around t. The helicity is strictly zero and $\pm 1$ for linearly and circularly polarized dipoles, respectively, and ranges between these values for elliptically polarized dipoles [18,48]. Eq. (3) confirms that two mechanisms dominate the lateral interaction forces induced on the particle: (i) the helicity of the polarization induced on the particle; and (ii) the local density of states provided by the surface, indirectly expressed in Eq. (3) through the imaginary part of the spatial derivative of the Green's function out of plane cross-terms at the particle position [48]. It should be noted that Eq. (3) is general and applies to any surface provided the adequate Green's function. For instance, it can be applied to analyze isotropic plasmonic surface, as done in Ref. [14,18]. There, the helicity of the particle polarization was exploited to convert the spin of incident circularly polarized light into lateral optical forces acting on the particles. In this work, we replace such surfaces with hyperbolic and extremely anisotropic metasurfaces able to significantly enhance the available density of states and, in turn, the strength of the induced lateral optical forces.

Fig. 2 illustrates the influence of the metasurface anisotropy on the lateral forces acting on a Rayleigh particle located nearby under circularly polarized plane wave illumination. According to Eq. (3), such anisotropy bounds the maximum force attainable over a metasurface. Specifically, Fig. 2a shows the lateral force $\mathbf{F}_\rho$ induced on the particle placed over a lossless MTS with $\sigma_{xx} = i10$ mS (i.e., metallic response, similar to $\text{Re}[\varepsilon_{xx}]<0$ in a bulk material) versus the conductivity along the y direction, $\sigma_{yy}$, thus analyzing all possible elliptical and hyperbolic topologies of the MTSs and going through its topological transition [30,32]. Fig. 2b depicts a similar analysis but detailing the lateral radiation pressure ($\mathbf{F}_\rho^0$) and interaction ($\mathbf{F}_\rho^s$) force that compose $\mathbf{F}_\rho$. Results, normalized with respect to the free-space scattering force $\mathbf{F}_0$ [48], confirm a very large enhancement of the induced forces due to the strongly dominant response of the interaction force over HMTSs (i.e., when $\text{sign}\{\text{Im}[\sigma_{xx}]\} \neq \text{sign}\{\text{Im}[\sigma_{yy}]\}$). Note that these metasurfaces support surface modes with high wavenumbers (see Fig. 2c) that are directionally excited by the scattering from the polarized particle thanks to the spin-Hall effect, providing a giant recoil force. Maximum enhancement is found near the topological transition of the metasurface ($\text{Im}[\sigma_{yy}] \approx 0$), which is a topology known to maximize the density of states [32]. However, the induced optical forces decrease for elliptical metasurfaces ($\text{sign}\{\text{Im}[\sigma_{xx}]\} = \text{sign}\{\text{Im}[\sigma_{yy}]\}$) due to the reduced density of states they provide and the limited wavenumber of the supported surface modes (see Fig. 2d). Fig. 2e completes this study exploring the lateral forces that appear over an anisotropic metasurface whose conductivity tensor components are simultaneously varied. The second and fourth quadrants of the figure show the response of HMTSs, confirming enhancements larger than 6 orders of magnitude over the free space scattering force. Importantly, such enhancement is robust against moderate deviations in the metasurface anisotropy, thus alleviating potential fabrication challenges when realizing specific designs. The first quadrant of Fig. 2e corresponds to elliptical metasurfaces that support transverse magnetic (TM) plasmons ($\text{Im}[\sigma_{yy}] > 0$ and $\text{Im}[\sigma_{xx}] > 0$) and provides significantly weaker optical forces than HMTSs. In bulk media, similar plasmons are supported by materials whose real part of their relatively permittivity is negative [13]. It should be emphasized that as the conductivity along one axis decreases, the induced forces increases due to the additional density of states provided by the structure and the larger wavenumbers of the supported SPPs. The limiting case appears again at the metasurface topological transition, where the induced optical forces reach their maximum value. Finally, the metasurfaces engineered in the third quadrant of Fig. 2e support elliptical transverse electric (TE) surface modes barely bounded to the surface and unable to induce significant optical forces on particles located nearby. Towards the normal direction, the acting force $\mathbf{F_z}$ may become either attractive or repulsive as a function of the distance between the particle and the metasurface [48], following similar trends as recently reported in the literature [49,50].

The second major mechanism that governs interaction forces over metasurfaces is the in-plane helicity of the polarization acquired by the particle. Specifically, the polarization spin around the x (y) axis induces optical forces towards the orthogonal axis within the plane, y (x), with a direction determined by the rotation handedness (see Eq. (3)). The maximum strength of the induced forces appears when the helicity is $\pm 1$, corresponding to circularly polarized dipoles. In practice, the helicity depends on the direction and polarization state of the incoming plane wave, the anisotropy of the MTSs, and the distance between the particle and the structure. A comprehensive evaluation of the influence of these parameters on the induced forces can be found in [48]. As happens in the case of standard plasmonic surfaces [18], the polarization of the incident light plays a key role to tailor the helicity and to control the strength of the forces. To illustrate this behavior, Fig. 3a shows the dipole polarization helicity around the x axis, $\eta_x$, versus the polarization state of the incoming light and the MTS conductivity value along the y axis, keeping constant the surface conductivity on the orthogonal axis ($\sigma_{xx}$). For the sake of simplicity, the incident light is aligned in this example with the x axis ($\phi_i = 0$), thus avoiding cross-polarization components on the reflected waves. The resulting helicity $\eta_x$ is zero when the total electric field does not simultaneously exhibit components along the y and z axis to induce a spin around x, as happens in this case with incident TE and TM waves. On the contrary, plane waves with elliptical and right/left handed circular polarization (RHCP/LHCP) possess these field components and thus might polarize the particle with a desired spin. Interestingly, the helicity increases as $|\sigma_{yy}|$ gets closer to zero because the MTS becomes almost transparent there and the particle acquires the polarization of the incident light. As $|\sigma_{yy}|$ increases, the fields reflected from the structure interfere with the incident wave, thus altering the particle polarization. In such cases, the helicity can be enhanced by simply changing the angle of incidence of the incoming light [48]. Note that the helicity around the y axis is only affected by the TM component of the incident light and therefore is constant $(\eta_y \approx -0.71)$ for all incident polarizations, except for purely TE waves, where it is strictly zero. Fig. 3b shows the lateral optical forces induced on the particle versus the polarization of the incoming light. As expected, the induced lateral forces are minimum for incident TE waves $(|\mathbf{F}_\rho^S| = 0, |\mathbf{F}_\rho| = |\mathbf{F}_\rho^0|)$, whereas their maximum strength depends on the interplay between the polarization states and the MTS anisotropy. Overall, plane waves with specific combination of polarization state and angle of incidence should be used to adequately polarize the particle and maximize the strength of the induced optical forces.

Hyperbolic and extremely anisotropic metasurfaces operating in the visible spectrum –where lasers with high output power are available– have been experimentally realized using single-crystal silver nanostructures [29]. Here, we numerically investigate the lateral optical forces acting on a electric Rayleigh particle located over such structure when it is illuminated by a plane wave with RHCP polarization, as illustrated in Fig. 4a. Fig. 4b depicts the strength of the induced forces versus frequency and compares it to the one obtained when the HMTS is replaced with bulk silver [51]. Numerical results obtained using COMSOL Multiphysics [48,52] confirm that the hyperbolic structure induces lateral optical forces with a strength more than three orders of magnitude larger than silver over a broadband frequency range (from 500 to 750 THz). Similar responses can be obtained using light with different polarization states and coming from other directions [48]. Fig. 4c-e illustrate the absolute magnitude of the Poynting vector of the fields scattered by the particle plotted exactly on top of the metasurface at different operation frequencies, confirming the directional excitation of highly confined surface modes. Inspecting these power plots, it is easy to observe the evolution of the metasurface topology from σ-near-zero to hyperbolic as frequency increases. Remarkably, the strength of the reported forces is even larger (up to 11 times) than the one found over bulk silver at its plasmonic resonance – located in the near ultraviolet at ~890 THz [48]. Note that the silver nanostructure considered here is designed to operate at visible frequencies [29] and it does not behave as a homogeneous metasurface in that band. Besides, such strength is comparable to the one appearing on strongly chiral particles in the presence of evanescent fields [53].

In conclusion, we have reported a giant enhancement of lateral optical forces acting on electrically polarizable Rayleigh particles located near hyperbolic and extremely anisotropic metasurfaces. The enhancement is enabled by the interplay between the increased density of states provided by these structures and the in-plane helicity of the polarization acquired by the particle. Theoretical and numerical results confirm that the induced optical forces are broadband, robust against fabrication tolerances, and quite resilient to loss (see [48] for further details). Hyperbolic and extremely anisotropic metasurfaces can be realized in different frequency bands to provide boosted lateral optical forces, using, for instance, nanostructured silver in the visible or nanostructured graphene in the terahertz and infrared bands [48]. Our analysis is based on a semi-classical electromagnetic approach that neglects other potential source of forces such as fluctuation induced, thermal, or quantum [54-56]. However, the very large enhancement of the reported optical forces and the fact that they appear over a broad frequency band strongly suggest that this response is of fundamental nature. In addition, we expect that anisotropic metasurfaces will also greatly influence fluctuation-induced forces since they depend on the same scattering problem analyzed here. Our findings might have significant implications in practice. In particular, hyperbolic metasurfaces seem ideal candidates to construct highly-performing nano-optical tweezers to precisely manipulate and trap nanoparticles as well as assembling them through optical binding, allowing to significantly reduce the intensity of the required laser beams and to prevent damaging due to photoheating. Furthermore, the large density of states provided by these structures will boost phenomena such as SERS and photoluminescence, finding exciting applications in fields such as nanophotonics, bioengineering and biochemistry.

This work was supported by the National Science Foundation with Grant. No. ECCS-1808400 and a CAREER Grant No. ECCS-1749177.

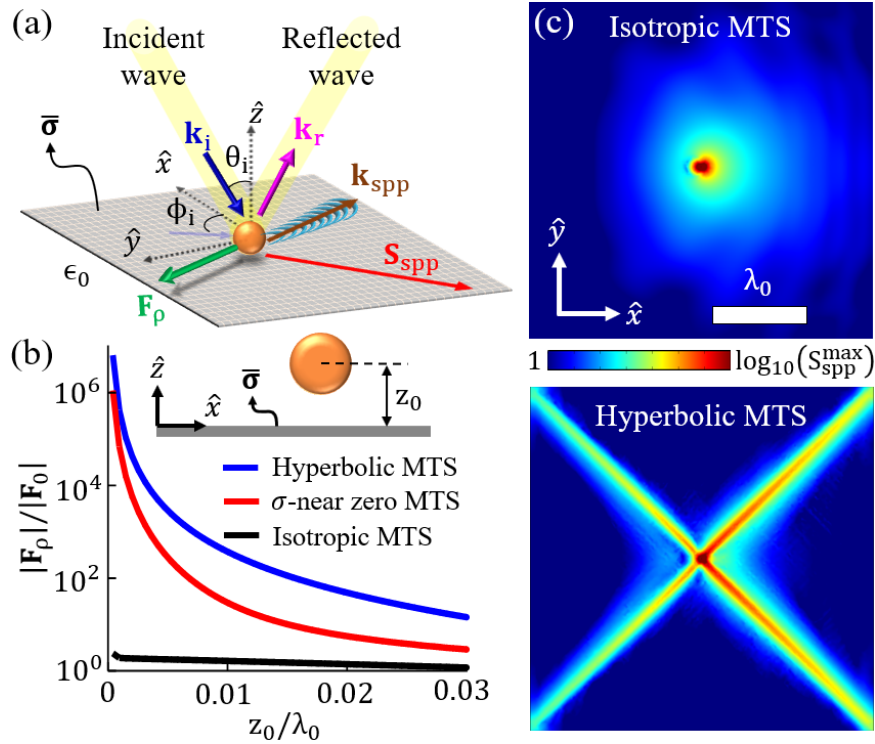

Fig. 1. Lateral optical forces induced on an electrically polarizable Rayleigh particle located in free space at a distance $z_0$ over an anisotropic metasurface characterized by a conductivity tensor $\bar{\sigma}$. (a) Schematic of the configuration. The particle scatters an incoming plane wave, exciting directional and confined surface plasmons –with wavevector $\mathbf{k_{spp}}$ and Poynting vector $\mathbf{S_{spp}}$– and experiencing a lateral recoil force $\mathbf{F_\rho}$. (b) Lateral optical forces induced on the particle versus its distance above isotropic (black line; with $\sigma_{xx} = \sigma_{yy} = i10\text{mS}$), $\sigma$-near zero (red line; $\sigma_{xx} = i10\text{mS}, \sigma_{yy} = -i0.1\text{mS}$), and hyperbolic (blue line; $\sigma_{xx} = i10\text{mS}, \sigma_{yy} = -i10\text{mS}$) metasurfaces. Results are normalized with respect to the free-space scattering force $\mathbf{F_0}$ [48]. (c) Absolute magnitude of the normalized Poynting vector on the isotropic (top) and hyperbolic (bottom) metasurfaces. The particle is located at $z_0$=1 um, has a radius r = 15 nm, relative permittivity $\varepsilon_p = 3$, and is illuminated by a TM plane wave at 8 THz coming from $\theta_i = 35^0$ and $\phi_i = 0^0$.

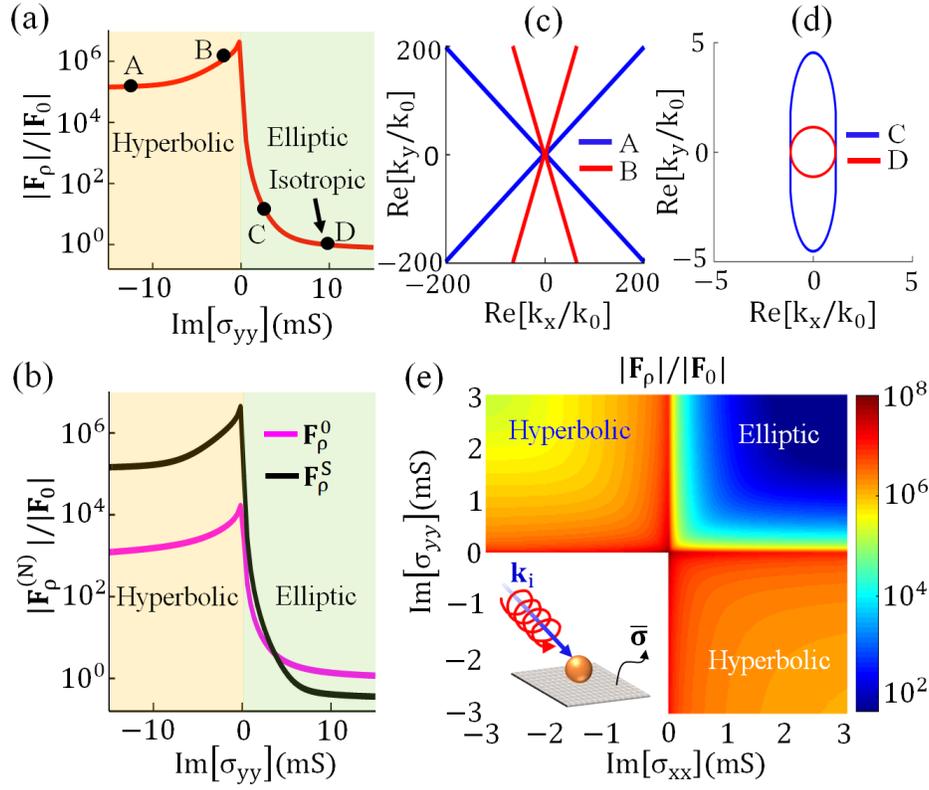

Fig. 2. Influence of the metasurface topology on the lateral optical forces induced on a Rayleigh particle located 40nm above the MTS and illuminated by a RHCP plane wave. (a) Forces induced on the particle versus the MTS conductivity along the y axis, $\sigma_{yy}$, for a fixed $\sigma_{xx} = i10$ mS. (b) Contributions to the lateral optical force: radiation pressure ($\mathbf{F}_\rho^0$, magenta) and interaction force ($\mathbf{F}_\rho^S$, black). (c), (d) Isofrequency contours of the A, B, C, and D metasurface topologies shown in panel (a). (e) Lateral forces versus the conductivity components of a lossless anisotropic metasurface. Results are normalized with respect to the free-space scattering force $\mathbf{F_0}$ [48]. Other parameters are as in Fig. 1.

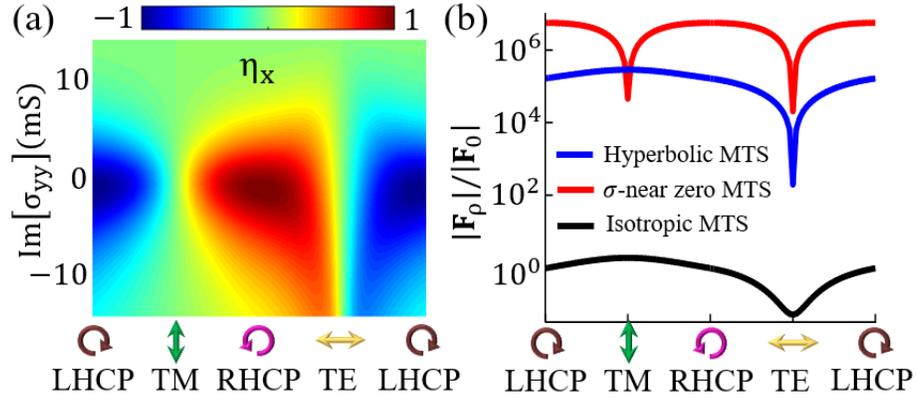

Fig. 3. Influence of the polarization of an incident plane wave on the lateral optical forces induced on a Rayleigh particle located 40nm above a lossless anisotropic metasurface. (a) Particle polarization helicity with respect to the x direction (see Fig. 1a). Results are plotted versus the MTS conductivity along the y axis, $\sigma_{yy}$, for a fixed $\sigma_{xx} = i10$ mS. (b) Lateral forces induced on the particle when it is located above isotropic (black; with $\sigma_{xx} = \sigma_{yy} = i10$mS), $\sigma$-near zero (red; $\sigma_{xx} = i10$mS, $\sigma_{yy} = -i0.1$mS), and hyperbolic (blue; $\sigma_{xx} = i10$mS, $\sigma_{yy} = -i10$mS) metasurfaces. The incoming plane wave has an axial ratio equal to 1 and polarization states that follows a circular trajectory on the Poincaré sphere. Other parameters are as in Fig. 1.

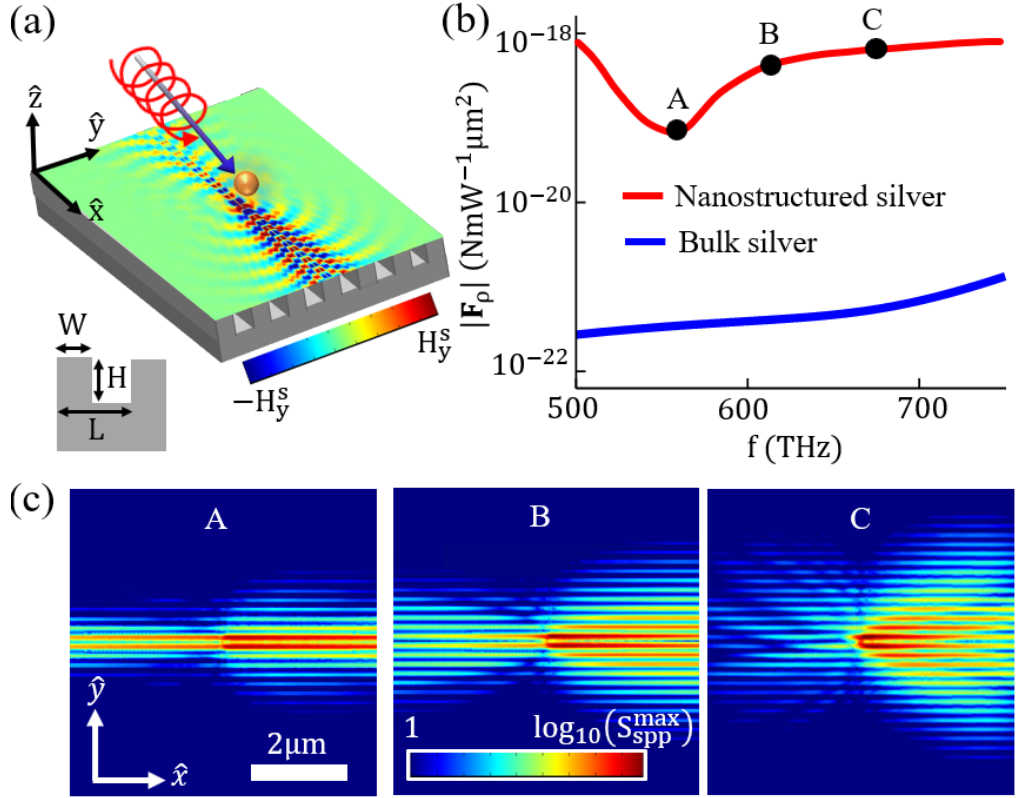

Fig. 4. Lateral optical forces on a Rayleigh particle located 25 nm above nanostructured silver [29]. (a) 3D schematic, showing an incident RHCP plane wave coming from $\theta_i = 35^0$ and $\phi_i = 0^0$. Superimposed field plot illustrates the y-component of the magnetic field excited on the structure (not to scale) at 612 THz due to the scattering process. (b) Lateral optical forces induced on the particle versus frequency. Results obtained when the particle is located above bulk silver [51] are included for comparison purposes. (c) Absolute magnitude of the normalized Poynting vector on the metasurface at the operation frequencies shown in panel (b). The particle has a radius r=15nm and relative permittivity $\varepsilon_p = 3$. The metasurface dimensions are W = 120 nm, H=80 nm, and L = 180 nm.